%% file: main.tex
\documentclass[final,3p,times]{elsarticle}
\usepackage{graphicx}
\usepackage{amsmath}
\usepackage{natbib}
\usepackage{subcaption}
\usepackage{tikz}
\usepackage[normalem]{ulem}
\usepackage[shortlabels,inline]{enumitem}
\usepackage{tabularx}
\usepackage{bm}
\usepackage{bbm}
\usepackage{xcolor}

\newcommand\Rey{\mbox{Re}}           

\colorlet{revision}{black}

\journal{Journal of Computational Physics}

\begin{document}

\begin{frontmatter}

\title{A High-Fidelity Methodology for Particle-Resolved Direct Numerical Simulations}
\author[aff1]{M. Houssem Kasbaoui\corref{cor1}}
\author[aff1]{Marcus Herrmann}
\cortext[cor1]{Corresponding author, email: houssem.kasbaoui@asu.edu}

\begin{abstract}
  \input{sections/abstract}
\end{abstract}


\end{frontmatter}

\input{body.tex}

\bibliography{references}
\end{document}

%% file: sections/abstract.tex
We present a novel computational method for direct numerical simulations of particle-laden flows with fully-resolved particles (PR-DNS). The method is based on the recently developed Volume-Filtering Immersed Boundary method [Dave et al, \emph{Journal of Computational Physics}, 487:112136, 2023] derived by volume-filtering the transport equations. This approach is mathematically and physically rigorous, in contrast to other PR-DNS methods which rely on ad-hoc numerical schemes to impose no-slip boundary conditions on the surface of particles. With the present PR-DNS strategy, we show that the ratio of filter size to particle diameter acts as a parameter that controls the level of fidelity. In the limit where this ratio is very small, a well-resolved PR-DNS is obtained. Conversely, when the ratio of filter size to particle diameter is large, a classic point-particle method is obtained. The discretization of the filtered equations is discussed and compared to other PR-DNS strategies based on direct-forcing immersed boundary methods. Numerical examples with sedimenting resolved particles are discussed.

%% file: body.tex
\input{sections/introduction.tex}
\input{sections/governing_equations.tex}

\input{sections/implementation.tex}

\section{Numerical examples}\label{sec:examples}
\input{sections/role_filter.tex}
\input{sections/unbounded_settling.tex}

\input{sections/low_density.tex}
\input{sections/confined_settling.tex}

\input{sections/conclusion.tex}

\section*{Acknowledgement}
The authors acknowledge support from the US National Science Foundation (award \#2028617, CBET-FD). Computing
resources were provided by Research Computing at Arizona State University.

%% file: sections/introduction.tex
\section{Introduction}
\label{sec:intro}

Expanding computing power offers the opportunity to perform simulations of particle-laden flows where the flow around individual particles is fully resolved. Such simulations, referred to hereafter as Particle-Resolved Direct Numerical simulations (PR-DNS) make it possible to investigate the micro-scale dynamics of particle-laden flows with the highest possible fidelity. However, the majority of previously proposed methods rely on ad-hoc numerical schemes that leave several questions unanswered.

\citet{uhlmannImmersedBoundaryMethod2005} was amongst the first to develop a methodology for PR-DNS that is both stable for a wide range of parameters and computationally efficient. In his approach, fast and scalable Cartesian grid solvers are employed in conjunction with an Immersed Boundary (IB) method to handle moving spheres. The IB method of \citet{uhlmannImmersedBoundaryMethod2005} is an extension of the original method of Peskin \citep{peskinImmersedBoundaryMethod2002}  wherein no-slip boundary conditions are imposed using ad-hoc forcing terms added to the right-hand side of the governing equations. This is a popular approach that has seen many variations \citep{yusofInteractionMassiveParticles1996,fadlunCombinedImmersedBoundaryFiniteDifference2000,kasbaouiDirectNumericalSimulations2021,verziccoImmersedBoundaryMethods2023}. \citet{uhlmannImmersedBoundaryMethod2005} builds the IB forcing term by requiring that the no-slip boundary condition is satisfied on Lagrangian points placed on the surface of the particles. A spreading operation is then performed by convolving with a regularized Dirac delta \citep{romaAdaptiveVersionImmersed1999} to turn the Lagrangian forcing into smooth Eulerian forcing. 
\citet{uhlmannImmersedBoundaryMethod2005} showed that this strategy can capture reasonably well the drag force on single and several spheres, fixed or moving, at Reynolds numbers up to $\sim 350$ with only 12 grid points across the diameter. 

However, there are several limitations the above mentioned IB method, the chief of which are (i) the ambiguity related to the placement and volumes of the Lagrangian forcing points and (ii) the unstable computation of hydrodynamic forces for low density ratios. These two problems stem from the ad-hocness of the method, i.e., the fact that the IB method is based purely on numerical considerations to impose boundary conditions without being derived from first principles. As a consequence, the volumes associated with the Lagrangian forcing points are not uniquely defined and become a tuning parameter. It is also not clear how these volumes are related to the width of the regularized Dirac delta, and what effect the latter has on the flow in the vicinity of the interface. Further, the method does not give a definitive answer to whether the forcing should apply throughout the particle volumes or not. \citet{uhlmannImmersedBoundaryMethod2005} constrains the forcing to the particle surfaces, and gives a formula to compute the hydrodynamic force on the particle that makes use of the acceleration of the internal flow. As acknowledged by \citet{uhlmannImmersedBoundaryMethod2005}, this approach is unstable for particle-to-fluid density ratios $\rho_p/\rho_f<1.2$.

Subsequent approaches to PR-DNS relied on variations of Uhlmann's method to improve drag predictions and/or stabilize the method for low density ratios. Because Uhlmann's method tends to over-predict drag, \citet{breugemSecondorderAccurateImmersed2012} proposed to retract inward the Lagrangian forcing points by a small distance, a fraction of the grid spacing $\Delta x$. In the view of \citet{breugemSecondorderAccurateImmersed2012}, retracting the Lagrangian forcing points is warranted because spreading the Lagrangian forcing onto the Eulerian grid with a numerical Dirac delta that has finite width leads to an apparent diameter that is larger than the actual one. However, the retraction distance depends on the particle Reynolds number and specifics of the numerical method \citep{tangMethodologyHighlyAccurate2014,luoImprovedDirectforcingImmersed2019}, which adds another tuning parameter. Instead of displacing the Lagrangian points, \citet{zhouAnalysisSpatiotemporalResolution2021} proposed to vary the associated volumes. \citet{zhouAnalysisSpatiotemporalResolution2021} view these volumes as relaxation weights that control how fast velocity errors decay. They propose to choose the largest permissible weights that keep the method stable. They carry out a stability calculation for a single fixed sphere, but extensions to multiple moving spheres are left out due to increased complexity and lack of universality. The lack of robustness of Uhlmann's method for low density ratios is a particularly thorny issue, as it prevents investigations of an important class of flows laden with neutrally buoyant particles. To address this problem, \citet{kempeImprovedImmersedBoundary2012} still used the formula provided by \citet{uhlmannImmersedBoundaryMethod2005}, but, instead of assuming rigid-body motion to simplify the calculation of the acceleration of the internal flow, they compute this term by direct integration. This approach requires an expensive numerical procedure to build a level set function that is $<0$ inside and $>0$ outside the particles. Later, \citet{tschisgaleNoniterativeImmersedBoundary2017} improved on \citep{kempeImprovedImmersedBoundary2012} by adding a semi-implicit time scheme for the particle motion which enable stable integrations with larger time steps.

In this work, we present a novel PR-DNS method that is based on the recently developed Volume-Filtering Immersed Boundary (VFIB) method \citep{daveVolumefilteringImmersedBoundary2023}. This method is derived from first principles by applying a volume-filter \citep{andersonFluidMechanicalDescription1967} to the Navier-Stokes equations first, then, discretizing the resulting filtered equations. Several unclosed terms arise from the filtering procedure, but as discussed in \citep{daveVolumefilteringImmersedBoundary2023} most of these terms drop in the DNS limit, i.e., when the filter size $\delta_f$ is much smaller than the particle size $d_p$ and any other characteristic length scale of the flow. In this limit, the no-slip boundary condition is used as closure in the interface force density representing the momentum exchange  between solid and fluid. Because this method is based on a strict physical and mathematical derivation, it does not suffer from issues related to ad-hocness as with methods based on Uhlmann's method. There are no problems with placing and sizing Lagrangian forcing points since no such concept is used in the derivation. In the VFIB method, the motion (or lack of) of the immersed solid is enforced with a force density that represents the momentum exchange between the two phases. This forcing expresses in terms of a surface integral on the immersed body. The discretization of this integral on a tessellation of the interface may be interpreted as Lagrangian forcing at the centroids of the elements making up the tessellation. Under this interpretation, the forcing points have volumes  that express in a closed form, in terms of the element surface area and surface density, in contrast with the ambiguous volumes in the method of \citet{uhlmannImmersedBoundaryMethod2005}. Further, we have shown in \citep{daveVolumefilteringImmersedBoundary2023} that the hydrodynamic stresses on an immersed solid have two contributions: resolved stresses, due to flow features larger than the filter size, and residual stresses. We have shown that it is imperative to include residual stresses and gave a formula to do so which incorporates the no-slip boundary condition as closure. With this approach, there is no need to integrate the acceleration of the internal flow.

The paper is organized as follows. We present the governing equations solved in the present PR-DNS strategy in section \ref{sec:gov_eq}. We review the derivation with volume-filtering, discuss the origin of resolved and residual stresses, and geometrical properties of the interface, namely, the interfacial volume fraction and the smearing length. In section \ref{sec:implementation}, we present the discretization and numerical implementation of these equations. We give several numerical examples in section \ref{sec:examples}, where we discuss the relative contributions of the resolved and residual stresses, provide comparisons with experimental data, and demonstrate the robustness and stability of the method for low density ratio and neutrally buoyant particles. Finally, we give concluding remarks in section \ref{sec:conclusion}.

%% file: sections/governing_equations.tex
\section{Governing equations}\label{sec:gov_eq}
\subsection{Equations summary}\label{ssec:summary}
We consider $N$ particles enclosed by surfaces $S_p$, for $p=1\dots N$, suspended in an incompressible fluid with constant density $\rho_f$ and viscosity $\mu_f$. The equations solved in this PR-DNS strategy are based on the volume-filtering framework established in \citep{daveVolumefilteringImmersedBoundary2023}. Here, we present a condensed summary of the equations for the particle and fluid phases. Sections \ref{ssec:fluid_phase} -- \ref{ssec:SDF} provide the specific details of the derivation.

Starting with the particle phase, we evolve each particle ``$p$'' by solving the following equations of motion
\begin{eqnarray}
  \frac{d\bm{x}_p}{dt}&=&\bm{u}_p \label{eq:part_1}\\
  m_p\frac{d\bm{u}_p}{dt} &=& \iint_{S_p}\bm{n}\cdot\bm{\tau}_fdS + \rho_p V_p\bm{g} \label{eq:part_2}\\
  I_p\frac{d\bm{\omega}_p}{dt} &=& \iint_{\bm{y}\in S_p}(\bm{y}-\bm{x}_p)\times\bm{n}\cdot\bm{\tau}_fdS
  \label{eq:part_3}
\end{eqnarray}
where $m_p$, $V_p$, $I_p$, $\bm{x}_p$, $\bm{u}_p$, and $\bm{\omega}_p$ denote the particle mass, volume, moment of inertia, centroid, linear velocity,  and angular velocity, respectively. In the above, $\bm{\tau}_f$ represents the unfiltered stress tensor due to the fluid surrounding the particle. It contains contributions from resolved and residual stresses. Section \ref{ssec:particle_phase} provides all the details on how to derive, close, and compute $\bm{\tau}_f$.

For the carrier phase, we solve volume-filtered equations for a mixture fluid with velocity $\bm{u}$ and pressure $p$. As discussed in \citep{daveVolumefilteringImmersedBoundary2023}, we obtain the mixture equations by introducing additional constitutive equations inside the solid and solving for the mixture mass and momentum. For reasons of computational efficiency, we assume fluid constitutive equations inside the particles, with identical density and viscosity as the fluid outside (see details in \S\ref{ssec:fluid_phase}). The resulting mixture equations for the conservation of mass and momentum are
\begin{eqnarray}
\nabla\cdot\bm{u}&=&0,\label{eq:vfib_1}\\
  \rho_f\left(\frac{\partial\bm{u}}{\partial t}+\nabla\cdot(\bm{u}\,\bm{u})\right)&=&-\nabla p+\mu_f\nabla^2\bm{u} + \rho_f \bm{g}\nonumber\\
  &+& \sum_p \iint_{\bm{y}\in S_p} \ell(\bm{y}) \left.\left( \rho_f\frac{d}{dt}\bm{u}^p_{I}  + \nabla p -\mu_f\nabla^2 \bm{u} - \rho_f \bm{g} \right)\right|_{\bm{y}}\mathcal{G}(\bm{x}-\bm{y})dS.\label{eq:vfib_2}
\end{eqnarray}
where $\mathcal{G}$ is a filter kernel of size $\delta_f$, $\bm{u}_I^p$ is the velocity at the solid-fluid interface of particle ``$p$'', $\bm{g}$ is gravity, and $\ell$ is the so-called smearing length. The latter quantity is a geometric property that depends on the filter kernel shape and size, as well as the local curvature of the interface (see \S\ref{ssec:SDF} for details).

There are many possible choices for the filter kernel $\mathcal{G}$ provided that it verifies three fundamental properties:
\begin{enumerate*}[(i)]
  \item $\mathcal{G}$ must be unitary, i.e, it integrates to one,
  \item symmetric, i.e., $\mathcal{G}(-\bm{y})=\mathcal{G}(\bm{y})$, and
  \item compact, i.e., $\mathcal{G}(\bm{y})=0$ for $|\bm{y}|>\delta_f$.
\end{enumerate*}  As we have shown in  \citep{daveVolumefilteringImmersedBoundary2023}, the choice of $\mathcal{G}$ matters little when the immersed surface is well-resolved by the filter. This is the case when  $\delta_f/d_p\ll 1$ for particles with characteristic size $d_p$. For this reason, we choose a simple filter kernel that is the product of three one-dimensional triangle kernels, i.e,
\begin{eqnarray}
  \mathcal{G}(x,y,z) &=& \mathcal{G}_1(x)\mathcal{G}_1(y)\mathcal{G}_1(z) \label{eq:filter_1}\\
  \mathcal{G}_1(r) &=& \left\{
  \begin{array}{ll}
    \frac{2}{\delta_f} \left(1-2\frac{|r|}{\delta_f}\right) & \text{if }|r|\leq\delta_f/2\\
    0 & \text{otherwise.}
  \end{array}\right.\label{eq:filter_2}
\end{eqnarray}

\subsection{Volume-filtering framework}\label{ssec:fluid_phase}
The governing equations (\ref{eq:vfib_1}) and (\ref{eq:vfib_2}) result from volume-filtering conservation equations inside and outside the solids. The derivation is presented in great detail in \citep{daveVolumefilteringImmersedBoundary2023}. For the sake of brevity and to help with subsequent discussion, we present only the highlights of the derivation.

The velocity $\bm{u}$ and pressure $p$ introduced in \S\ref{ssec:summary} are mixture quantities defined as,
\begin{eqnarray*}
	\bm{u}&=&\alpha_f\overline{\bm{u}}_f+\alpha_s\overline{\bm{u}}_s,\\
	p	  &=&\alpha_f\overline{p}_f+\alpha_s\overline{p}_s
\end{eqnarray*}
where 
$\alpha_f$, $\overline{\bm{u}}_f$, and  $\overline{p}_f$ are the fluid volume fraction, filtered velocity, and filtered pressure, respectively. Similarly, $\alpha_s$, $\overline{\bm{u}}_s$, and  $\overline{p}_s$ denote the solid quantities. Volume-filtering is carried out using a kernel $\mathcal{G}$ that is (i) unitary, (ii) symmetric, and (iii) compact with size $\delta_f$ \citep{daveVolumefilteringImmersedBoundary2023}. The fluid and solid volume fractions are obtained by filtering the phase-indicator functions $\mathbbm{1}_f$ and $\mathbbm{1}_s$, 
\begin{eqnarray}
	\alpha_f(\bm{x})&=&\iiint_{\bm{y}\in \mathbbm{R}^3} \mathbbm{1}_f(\bm{y})\mathcal{G}(\bm{x}-\bm{y}) dV\label{eq:vol_frac}\\
	\alpha_s(\bm{x})&=&\iiint_{\bm{y}\in \mathbbm{R}^3} \mathbbm{1}_s(\bm{y})\mathcal{G}(\bm{x}-\bm{y}) dV
\end{eqnarray}
Here, $\alpha_f(\bm{x})$ and $\alpha_s(\bm{x})$ represent the fraction of space occupied by the fluid or solid within a region of size $\delta_f$ centered at an arbitrary position $\bm{x}$. Within the fluid region sufficiently far away from the interface such that the solid region does not intersect the filter kernel, $\alpha_f=1$ and $\alpha_s=0$.  Conversely, $\alpha_f=0$ and $\alpha_s=1$ within the solid region and sufficiently away from the interface. Note that the two volume fractions are constrained by $\alpha_f+\alpha_s=1$. Further, because of the finite width $\delta_f$ of the kernel $\mathcal{G}$, the volume fractions at positions within $\delta_f/2$ from the interface take values between 0 and 1. Thus, $\delta_f$ represents the sharpness or the filter kernel and controls the resolution of the method. 

The filtered velocities are defined from the unfiltered quantities, also called \emph{point-wise} quantities \citep{andersonFluidMechanicalDescription1967,jacksonDynamicsFluidizedParticles2000}, as follows
 \begin{eqnarray}
	\alpha_f(\bm{x})\overline{\bm{u}}_f(\bm{x})&=&\iiint_{\bm{y}\in \mathbbm{R}^3} \mathbbm{1}_f(\bm{y})\bm{u}_f(\bm{y})\mathcal{G}(\bm{x}-\bm{y}) dV,\\
	\alpha_s(\bm{x})\overline{\bm{u}}_s(\bm{x})&=&\iiint_{\bm{y}\in \mathbbm{R}^3} \mathbbm{1}_s(\bm{y})\bm{u}_s(\bm{y})\mathcal{G}(\bm{x}-\bm{y}) dV.
\end{eqnarray}
As with the volume fractions, the filtered velocity of one phase decays to zero into the other phase within a distance $\delta_f/2$ from the interface.

 Within the fluid, the point-wise governing equations are the incompressible Navier-Stokes equations
 \begin{eqnarray}
\nabla\cdot\bm{u}_f&=&0\label{eq:gov_1},\\
  \rho_f\left(\frac{\partial\bm{u}_f}{\partial t}+\nabla\cdot(\bm{u}_f\,\bm{u}_f)\right)&=&-\nabla p_f+\mu_f\nabla^2\bm{u}_f\label{eq:gov_2}
\end{eqnarray}
For simplicity, we assume that the fluid density and viscosity are constant. These equations must be completed with appropriate boundary conditions, namely, the no-slip boundary condition on the surfaces of the particles,
\begin{equation}
	\bm{u}_f(\bm{x})=\bm{u}_I^p(\bm{x}) \text{ for }\bm{x}\in S_p\quad p= 1,\dots,N.
\end{equation}

Filtering the fluid equations (\ref{eq:gov_1}) and (\ref{eq:gov_2}) leads to the following equations,
\begin{eqnarray}
  \frac{\partial\alpha_f}{\partial t}+\nabla\cdot(\alpha_f\overline{\bm{u}}_f)&=&0, \label{eq:gov_filtered_1}\\
  \rho_f\left(\frac{\partial}{\partial t}(\alpha_f\overline{\bm{u}}_f)+\nabla\cdot(\alpha_f\overline{\bm{u}}_f\,\overline{\bm{u}}_f)\right)&=&\alpha_f\nabla\cdot\left(\overline{\bm{\tau}}_f+\bm{R}_\mu-\bm{\tau}_\mathrm{SFS}\right)+\bm{F}_I \label{eq:gov_filtered_2}
\label{eq:gov_4}
\end{eqnarray}
where $\overline{\bm{\tau}}_f=-\overline{p}_f\bm{I}+\mu_f\left(\nabla\overline{\bm{u}}_f+\nabla\overline{\bm{u}}_f^T-\frac{2}{3}(\nabla\cdot\overline{\bm{u}}_f)\bm{I}\right)$ is the filtered stress tensor, $\bm{\tau}_\mathrm{SFS}=\overline{\bm{u}_f\bm{u}_f}-\overline{\bm{u}}_f\overline{\bm{u}}_f$ is the subfilter scale tensor, and $\bm{R}_\mu$ is the residual viscous stress tensor \citep{daveVolumefilteringImmersedBoundary2023}. The term $\bm{F}_I$ represents the momentum exchange between the solid and fluid phases. It expresses as
\begin{equation}
	\bm{F}_I(\bm{x})=\sum_p\iint_{\bm{y}\in S_p} \bm{n}\cdot\left(\bm{\tau}_f-\overline{\bm{\tau}}_f-\bm{R}_\mu+\bm{\tau}_\mathrm{SFS}\right)\mathcal{G}(\bm{x}-\bm{y})dS
\end{equation}
Following \citep{daveVolumefilteringImmersedBoundary2023}, this term can also be expressed as
\begin{equation}
	\bm{F}_I(\bm{x})=\sum_p\iint_{\bm{y}\in S_p} \rho_f\alpha_{f,I}\ell\left(
	\frac{d}{dt}\left(\bm{u}^p_I+ \bm{u}^p_{s,I}\right) 
	- \nabla\cdot\left(\overline{\bm{\tau}}_f + \bm{R}_\mu -\bm{\tau}_\mathrm{SFS}\right)
	\right)\mathcal{G}(\bm{x}-\bm{y})dS
\end{equation}
where $\alpha_{f,I}$ is the fluid volume fraction at the interface and the term $\bm{u}^p_{s,I}=\overline{\bm{u}}_I-\bm{u}^p_I$ represents the slip at the interface between the filtered velocity and the no-slip boundary condition. 

\begin{figure} \centering
  \includegraphics[width=1\linewidth]{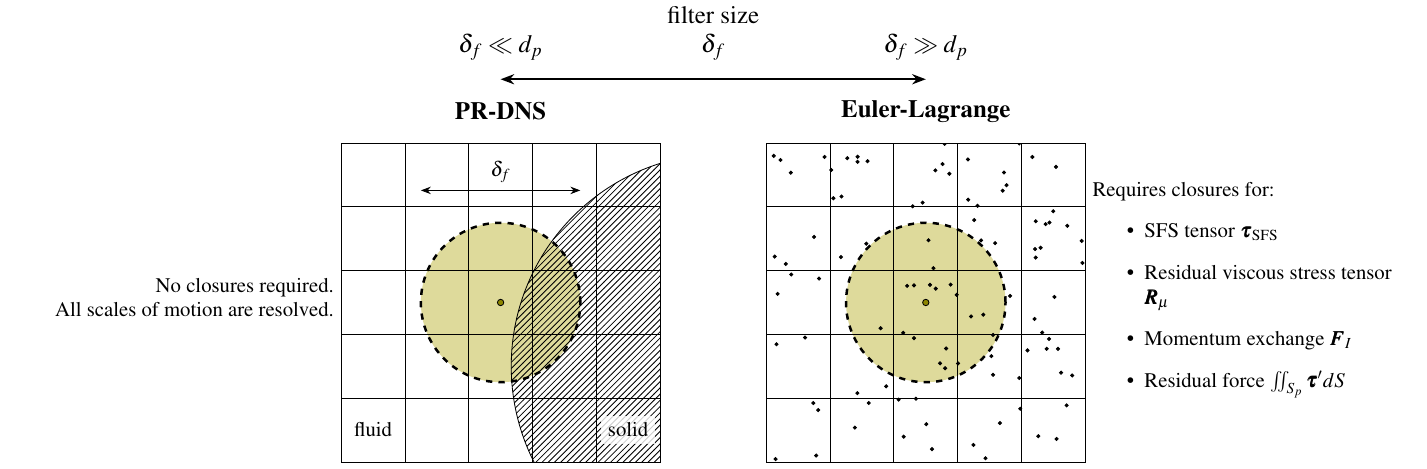}
  \caption{The ratio of filter size to particle diameter controls the resolution in the volume-filtering framework. No closures are needed in the PR-DNS limit ($\delta_f\ll d_p$) considered in this study.\label{fig:schematic}}
\end{figure}

Filtering introduces a closure problem (see schematic in figure \ref{fig:schematic}). In general, models for the interfacial slip $\bm{u}_{s,I}$, the subfilter scale tensor $\bm{\tau}_\mathrm{SFS}$, and  the residual viscous stress tensor $\bm{R}_\mu$ must be supplied. However, for well resolved particles, i.e, $\delta_f/d_p\ll 1$, these unclosed terms can be neglected. We refer to this as the DNS limit of the volume-filtering method, and neglect the unclosed terms in what follows.

To obtain governing equations for the mixture quantities, we must choose equations to solve inside the solid. The most straightforward choice is to assume rigid body motion or some other solid mechanics model. But, as we explain in \citep{daveVolumefilteringImmersedBoundary2023}, it is computationally more advantageous to assume that the particles are thin, rigid shells that enclose an internal fluid with density and viscosity that match those of the outside fluid. With this assumption, volume-filtered governing equations similar to equations (\ref{eq:gov_filtered_1}) and (\ref{eq:gov_filtered_2}) can be written for the internal fluid. Summing up the governing equations for the internal and external fluid yields the mixture equations (\ref{eq:vfib_1}) and (\ref{eq:vfib_2}) presented in the summary section \S\ref{ssec:summary}.

\subsection{Fluid stresses on resolved particles}\label{ssec:particle_phase}
Calculating the hydrodynamic force exerted on particle ``$p$'' requires integrating the fluid stress on the particle surface, i.e., 
\begin{equation}
	\bm{F}^p_h=\iint_{S_p}\bm{n}\cdot\bm{\tau}_fdS.
\end{equation}
In the expression above, $\bm{\tau}_f$ is the unfiltered fluid stress tensor. In the present volume-filtered formulation, $\bm{\tau}_f$ decomposes into filtered and residual parts: $\bm{\tau}_f=\overline{\bm{\tau}}_f+\bm{\tau}_f'$. In the DNS limit ($\delta_f/d_p\ll 1$), these two contributions can be calculated from the mixture quantities $\bm{u}$ and $p$ as follows \citep{daveVolumefilteringImmersedBoundary2023}
\begin{eqnarray}
  \bm{n}\cdot\overline{\bm{\tau}}_f&=&-p \bm{n}+\mu(\nabla\bm{u}+\nabla\bm{u}^T)\cdot\bm{n}\label{eq:part_4}\\
  \bm{n}\cdot\bm{\tau}_f'&=& \alpha_{f,I}\ell\left( \rho_f\frac{d}{dt}\bm{u}^p_{I}  + \nabla p -\mu_f\nabla^2 \bm{u} \right) \label{eq:part_5}
\end{eqnarray}
Hence, the hydrodynamic force on particle ``$p$'' has the following two contributions
\begin{equation}
	\bm{F}^p_h=\underbrace{\iint_{S_p}-p \bm{n}+\mu(\nabla\bm{u}+\nabla\bm{u}^T)\cdot\bm{n}dS}_{\overline{\bm{F}}_h^p}
	 + \underbrace{\iint_{S_p} \alpha_{f,I}\ell\left( \rho_f\frac{d}{dt}\bm{u}^p_{I}  + \nabla p -\mu_f\nabla^2 \bm{u} \right)dS}_{{\bm{F}^p_h}'}. \label{eq:part_6}
\end{equation}
where the first integral on the right-hand side of (\ref{eq:part_6}) gives the resolved force $\overline{\bm{F}}_h^p$, and the second integral gives the residual force ${\bm{F}^p_h}'$.

To understand the significance of the resolved and residual forces, it is useful to consider the limit where the filter size is much larger than the particle diameter. This is the usual limit considered when deriving Euler-Lagrange models \citep{capecelatroEulerLagrangeStrategy2013,irelandImprovingParticleDrag2017}. With the assumption $\delta_f/d_p\gg 1$, the resolved force is usually approximated as
\begin{eqnarray}
	\overline{\bm{F}}^p_h&=& \iint_{S_p} \bm{n}\cdot\overline{\bm{\tau}}_f dS
	 =     \iiint_{V_p}\nabla\cdot \overline{\bm{\tau}}_f dV 
   \simeq V_p\nabla\cdot \overline{\bm{\tau}}_f.
   \label{eq:resolved_force}
\end{eqnarray}
Thus, the resolved force may be interpreted as the force resulting from the undisturbed flow \citep{maxeyEquationMotionSmall1983,irelandImprovingParticleDrag2017}. The residual force cannot be evaluated in closed form when $\delta_f/d_p\gg 1$, as it involves unresolved fluctuations of the fluid stress tensor. Thus, ${\bm{F}^p_h}'$ is commonly modeled as a drag force. For example, Stokes drag would be used to model the residual force for spherical particles with Reynolds number $\Rey_p\ll 1$, 
\begin{equation}
	{\bm{F}^p_h}'\simeq m_p \frac{(\bm{u}_f(\bm{x}_p,t)- \bm{u}_p)}{\tau_p} \label{eq:EL_residual}
\end{equation}
where $\tau_p=\rho_pd_p^2/(18 \mu_f)$ is the particle response time. Depending on the flow regime, additional force models such as added mass, Saffman lift, and Basset History may be appended to the drag force in (\ref{eq:EL_residual}), e.g. \citep{kasbaouiTurbulenceModulationSettling2019,shuaiAcceleratedDecayLamb2022,daveMechanismsDragReduction2023,shuaiInstabilityDustyVortex2022,shuaiMergerCorotatingVortices2024,vandorenTurbulenceModulationDense2023}.

Unlike in Euler-Lagrange models, both filtered and residual stresses can be directly evaluated using equations (\ref{eq:part_4}) and (\ref{eq:part_5}) in the present PR-DNS framework. The no-slip boundary condition serves as closure for the residual force to obtain the closed form above. Note that the residual force does not vanish even in the limit of very high resolutions, i.e., $\delta_f/d_p$ very small. This is due to the fact that expressions (\ref{eq:part_4}) and (\ref{eq:part_5}) must be evaluated on the solid-fluid interface, which is within the transition region for the filtered quantities. In \S\ref{ssec:role_filter}, we show that for vanishingly small values of $\delta_f/d_p$, the residual and filtered forces have equal contributions. With increasing $\delta_f/d_p$, the residual force becomes larger than the filtered one.

\subsection{Volume fraction}\label{ssec:volume_fraction}

\begin{figure}\centering
	\includegraphics[width=.45\linewidth]{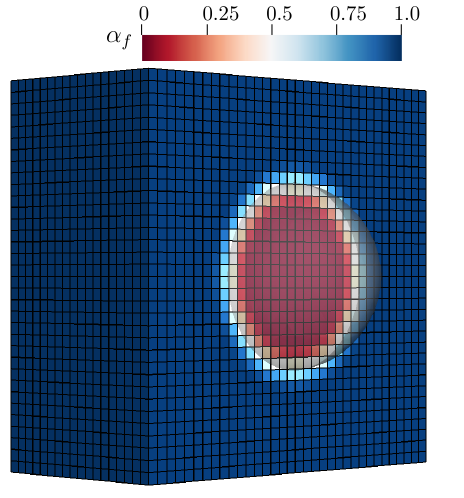}
	\caption{Fluid volume fraction field $\alpha_f$ in the vicinity of an isolated spherical particle obtained by solving equation (\ref{eq:volume_fraction}). In this example, the filter size is $ \delta_f=(1/4)d_p$. The grid is uniform with spacing $\Delta x/\delta_f=1/4$, which is sufficiently fine to resolve the transition region. The sphere is meshed with a Delaunay triangulation such that the approximate triangle size is $\Delta s=\delta_f/8$. The semi-transparent surface shows the iso-surface $\alpha_{f,I}=0.5+0.176\times(\delta_f/D)$, which represents the interfacial fluid volume fraction and the location of the solid-fluid interface.\label{fig:volfrac_single_sphere}}
\end{figure}
Although the residual stresses are given in the closed form (\ref{eq:part_6}), evaluating these terms requires calculating  the fluid volume fraction at the interface  $\alpha_{f,I}$. The definition (\ref{eq:vol_frac}) does not provide an efficient way of calculating the volume fraction field $\alpha_f$ because it requires computationally expensive procedures to build and integrate the fluid phase-indicator function $\mathbbm{1}_f$ \citep{apteNumericalMethodFully2009,breugemSecondorderAccurateImmersed2012}. Instead, $\alpha_f$ can be computed efficiently for arbitrarily shaped particles by solving the following Poisson equation \citep{unverdiFronttrackingMethodViscous1992,daveVolumefilteringImmersedBoundary2023},
\begin{eqnarray}
  \nabla^2\alpha_f &=& \nabla \cdot \left(\sum_p\iint_{\bm{y}\in S_p} \bm{n}(\bm{y})\mathcal{G}(\bm{x}-\bm{y}) dS\right). \label{eq:volume_fraction}
\end{eqnarray}

Figure \ref{fig:volfrac_single_sphere} shows the fluid volume fraction field $\alpha_f$ obtained by solving equation (\ref{eq:volume_fraction}) for an isolated spherical particle. Note that we deliberately choose a coarse filter size $\delta_f=(1/4)d_p$ to highlight the transition region around the interface. 
The integration of equation (\ref{eq:volume_fraction}) is performed with second order schemes and an algebraic multigrid method on a uniform Cartesian grid as described in \citep{daveVolumefilteringImmersedBoundary2023}. To provide sufficient resolution of the kernel and the transition region, we choose the grid spacing $\Delta x$ such that $\Delta x=\delta_f/4$. Thus, 16 grid points lie across the diameter for the case with $\delta_f=(1/4)d_p$ shown in figure \ref{fig:volfrac_single_sphere}. The surface of the sphere is discretized using a Delaunay triangulation with approximate element size $\Delta s=\delta_f/8$.  Although solving the Poisson equation (\ref{eq:volume_fraction}) increases the overall computational cost of the method,  it remains significantly faster than computing the volume fraction from the definition (\ref{eq:vol_frac}).

\begin{figure}\centering
   \includegraphics[width=.7\linewidth]{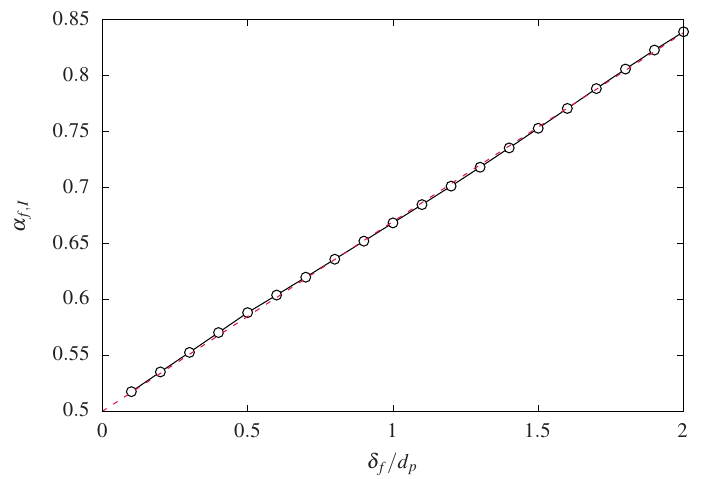}
  \caption{Variation of the interfacial fluid volume fraction with filter size for an isolated sphere. The red dashed line is the fit $\alpha_{f,I}=0.5+1.7\,10^{-1}\times(\delta_f/d_p)-3.85\,10^{-4}\times(\delta_f/d_p)^2$.\label{fig:interface_volume_fraction}}
\end{figure}
Solving equation (\ref{eq:volume_fraction}) gives the complete volume fraction field, both within and away from the particle. This serves two purposes: (i) computing the interfacial volume fraction $\alpha_{f,I}$ needed to evaluate the residual stresses in equation (\ref{eq:part_5}), and (ii) visualizing the location of particles using the iso-surface $\alpha_f=\alpha_{f,I}$.

Here, we emphasize that the interfacial fluid volume fraction $\alpha_{f,I}$ is equal to 0.5 only when the interface is perfectly planar.
For spherical particles, $\alpha_{f,I}$ depends on the relative size of the filter kernel $\delta_f$ and the particle diameter $d_p$. To illustrate this behavior, we show the variation of $\alpha_{f,I}$ for  $0.1\leq\delta_f/d_p\leq 2$ for an isolated spherical particle in figure \ref{fig:interface_volume_fraction}. We generate this data by solving (\ref{eq:volume_fraction}) on a fine grid $\Delta x=\min(d_p/12,\delta_f/4)$. We then use trilinear interpolations to evaluate $\alpha_f$ at locations on the particle surface.
Note that  $\alpha_{f,I}$ is the same at all location on the surface owing to spherical symmetry. As evidenced in figure \ref{fig:interface_volume_fraction}, the interfacial fluid volume fraction approaches 0.5 for very fine resolutions.
This is because the curvature of the sphere on the scale of the filter becomes increasingly negligible when $\delta_f/d_p\rightarrow 0$, making the interface appear flat. With increasing ratio $\delta_f/d_p$, the interfacial fluid volume fraction increases as the concave curvature of the sphere leads to comparatively less solid than fluid under the filter kernel. For $\delta_f/d_p<0.5$, the variation of $\alpha_{f,I}$ with  $\delta_f/d_p$ is linear and may be fitted by $\alpha_{f,I}=0.5+1.7\,10^{-1}\times(\delta_f/d_p)$. Note that this fit applies for the triangle kernel only. Although not shown here, other kernels yield different fits. 

Solving the Poisson equation (\ref{eq:volume_fraction}) may be forgone in unbounded and dilute particulate flows. In such flows, the particles appear isolated as particle-particle and particle-wall contact is negligibly rare. Thus, the fit $\alpha_{f,I}=0.5+0.176\times(\delta_f/d_p)$ can be directly injected in equation (\ref{eq:part_5}) to speed-up the calculation of the residual stresses without solving the volume fraction equation (\ref{eq:volume_fraction}).

\begin{figure}\centering
  \includegraphics[width=0.95\linewidth]{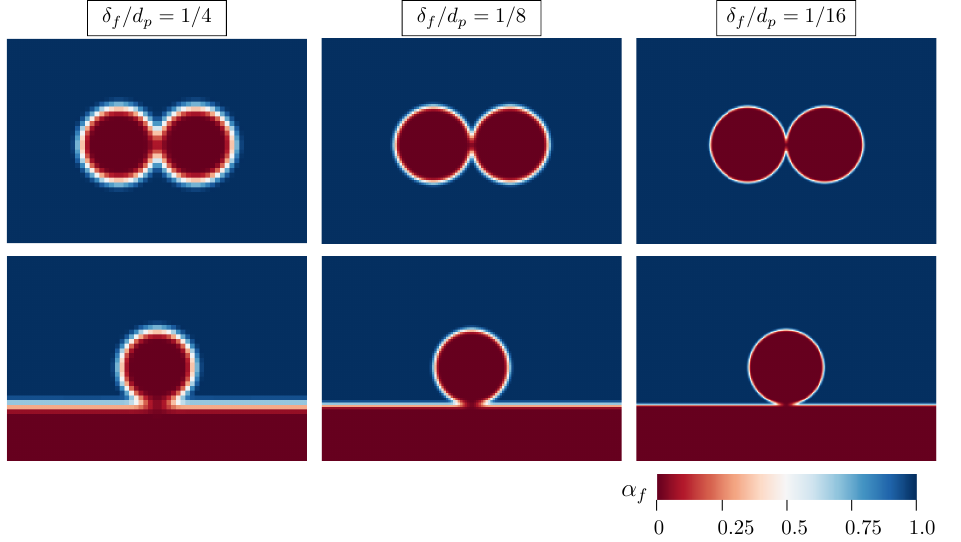}
  \caption{Fluid volume fraction field for two spheres at contact (top panels) and a sphere resting on a wall (bottom) at different interface resolutions. At the contact region, the interfacial fluid volume fraction $\alpha_{f,I}$ vanishes which leads to a local cancelation of the residual stresses on the particle. \label{fig:contact}}
\end{figure}

In contrast, configurations where particle-particle and particle-wall contact are encountered lead to non-trivial volume fraction fields. In these cases, the interfacial volume fraction $\alpha_{f,I}$ varies on the surface of the particle. Most notably,  $\alpha_{f,I}$ drops considerably at the particles-particle and particle-wall contact points. Figure \ref{fig:contact} illustrates this effect in the cases of a spherical particle contacting another spherical particle of equal diameter or a flat wall. The figure shows the fluid volume fraction field $\alpha_f$ for the filter sizes $\delta_f=(1/4)\times d_p$, $(1/8)\times d_p$, and $(1/16)\times d_p$. In all these cases, $\alpha_f$ reduces considerably at the contact point between solids. Consequently, the residual stresses in equation (\ref{eq:part_5}) from the contact region also reduce.

\subsection{The smearing length and surface density function}\label{ssec:SDF}
\begin{figure}\centering
	\includegraphics[width=.7\linewidth]{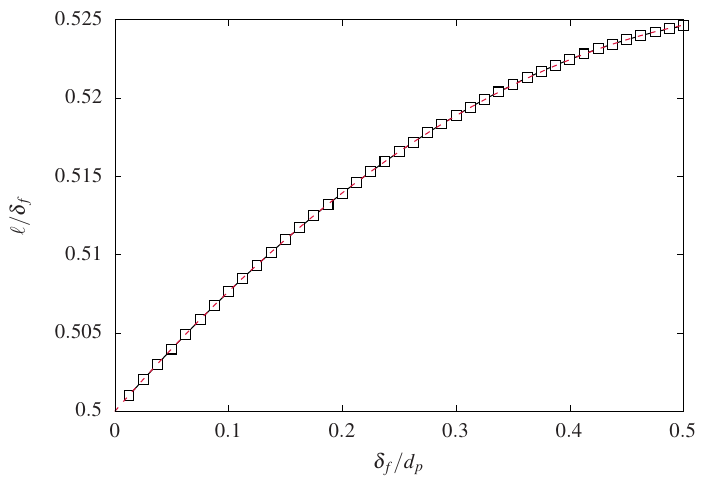}
	\caption{Variation of the smearing length with interface resolution for an isolated spherical particle. The red dashed line is the fit $\ell/\delta_f=0.5+8.33\,10^{-2}\times(\delta_f/d_p)-6.80\,10^{-2}\times(\delta_f/d_p)^2$.\label{fig:ell}}
\end{figure}

Similar to the interfacial volume fraction $\alpha_{f,I}$, the smearing length $\ell$ is another geometric property of the interface that depends on the local interface curvature, shape, and size of the filter kernel \citep{daveVolumefilteringImmersedBoundary2023}. It appears in the coupling term in the mixture equation (\ref{eq:vfib_2}) and in the residual stresses in (\ref{eq:part_5}). As such, it must be computed carefully to yield accurate results in this PR-DNS strategy.

The smearing length is tightly connected to the surface density function,
\begin{equation}
  \Sigma(\bm{x})=\sum_{p=1}^N \iint_{\bm{y}\in S_p}\mathcal{G}(\bm{y}-\bm{x})dS
\end{equation}
which has dimensions of inverse of length. Integrating $\Sigma$ over the entire domain gives the total interfacial area. In the case of $N$-monodisperse particles of diameter $d_p$, integrating $\Sigma$ gives $N\pi d_p^2$. 

The smearing length $\ell(\bm{y}_I)$ at a location $\bm{y}_I$ on the solid-fluid interface is the inverse of $\Sigma$ at that location, i.e., $\ell(\bm{y}_I)=(\Sigma(\bm{y}_I))^{-1}$. For a perfectly flat interface, $\ell$ can be easily evaluated as $\ell=\mathcal{G}(\bm{0})^{-1/3}$. Using the filter in (\ref{eq:filter_1}) and (\ref{eq:filter_2}), $\ell=\delta_f/2$ at locations where the interface is flat.

Figure \ref{fig:ell} shows the variation of the smearing length $\ell$ with the ratio $\delta_f/d_p$ for an isolated spherical particle with diameter $d_p$. With vanishing $\delta_f/d_p$, $\ell$ approaches the value for a flat interface $\mathcal{G}(\bm{0})^{-1/3}=\delta_f/2$ as the particle curvature becomes negligible on the scale of the filter kernel. For larger values of $\delta_f/d_p$ but less than 0.5, the fit $\ell/\delta_f=0.5+8.33\,10^{-2}\times(\delta_f/d_p)-6.80\,10^{-2}\times(\delta_f/d_p)^2$ provides a good approximation of the smearing length.

%% file: sections/implementation.tex
\section{Numerical implementation}\label{sec:implementation}

Having addressed the theoretical background in \S\ref{sec:gov_eq}, we now address the numerical implementation of the method. Our strategy relies on iterative Adams-Moulton and Crank-Nicolson schemes to update particle and fluid solutions from time stage $n$ to $n+1$, respectively. This a robust second-order procedure that we have used in \citep{daveVolumefilteringImmersedBoundary2023} to implement the volume-filtering immersed boundary method, and in \citep{kasbaouiAlgorithmSolvingNavier2017} and \citep{kasbaouiClusteringEulerEuler2019} to implement shear-periodic boundary conditions and Euler-Euler solvers for particle-laden flows, respectively. 

In the following, we describe the procedure to advance the solution from time stage $n$ to $n+1$ when accounting for the resolved particles:

\paragraph{Step 0} Before initiating any sub-iterations, we start by setting the quantities at sub-iteration $k=0$ and time level $n+1$ to the old values at time level $n$: 
  \begin{eqnarray*}
  	\bm{u}_0^{n+1}&=&\bm{u}^n					\quad\text{(fluid velocity)}\\
	p_0^{n+1}&=&p^n								\quad\text{(fluid pressure)}\\
  	\bm{x}_{p,0}^{n+1}&=&\bm{x}_p^{n}           \quad\text{(particle centroid position)}\\
  	\bm{u}_{p,0}^{n+1}&=&\bm{u}_p^{n}           \quad\text{(particle centroid velocity)}\\
  	\bm{\omega}_{p,0}^{n+1}&=&\bm{\omega}_p^{n} \quad\text{(particle centroid angular velocity)}
  \end{eqnarray*}
 
	Next, we increment the counter $k$ and initiate the sub-iterations.
	
\paragraph{Step 1 (Particle Solver)} In steps 1.1 to 1.5, we integrate forward the particle-phase equations (\ref{eq:part_1}) --  (\ref{eq:part_3}) and update the particle boundaries.

\textbf{Step 1.1} In this step, we determine the hydrodynamic force $\bm{F}_{h,p,k}^{n+1}$ and torque $\bm{T}_{h,p,k}^{n+1}$ applied on a particle ``$p$'' using data from the previous sub-iteration $k$ and time level $n+1$. Discretizing the integrals in equation (\ref{eq:part_6}) with the mid-point rule gives:
\begin{eqnarray}
	\bm{F}_{h,p,k}^{n+1}&=&\sum_{t\in S_p}\left(\bm{n}\cdot\bm{\tau}_{f}\right)^{n+1}_{t,k} A_t\\
	\bm{T}_{h,p,k}^{n+1}&=&\sum_{t}(\bm{x}_{t,k}^{n+1}-\bm{x}_{p,k}^{n+1})\times\left(\bm{n}\cdot\bm{\tau}_{f}\right)^{n+1}_{t,k} A_t
\end{eqnarray}
where the set $S_p$ represents the elements of the surface mesh of this particle, $\bm{x}_t$ represents the centroid of element $t$, $\bm{n}_t$ the normal pointing from the solid to the fluid at $\bm{x}_t$, and $A_t$ the associated surface area. Recall that the hydrodynamic stresses have resolved and subfilter components (\S\ref{ssec:particle_phase}), such that, 
\begin{eqnarray}
	\left(\bm{n}\cdot\bm{\tau}_{f}\right)^{n+1}_{t,k}&=& \left(\bm{n}\cdot\overline{\bm{\tau}}_{f}\right)^{n+1}_{t,k} + \left(\bm{n}\cdot\bm{\tau}'_{f}\right)^{n+1}_{t,k},\\
	\left(\bm{n}\cdot\overline{\bm{\tau}}_{f}\right)^{n+1}_{t,k}&=&-\left.p_{k}^{n+1}\right|_t\bm{n}_{t,k}^{n+1}+\mu \bm{n}_{t,k}^{n+1}\cdot\left.\left(\nabla \bm{u}_{k}^{n+1}+\left(\nabla \bm{u}_{k}^{n+1}\right)^T\right)\right|_t, \\
	\left(\bm{n}\cdot\bm{\tau}'_{f}\right)^{n+1}_{t,k}&=&\frac{\left.\alpha_{f,k}^{n+1}\right|_t}{\left.\Sigma_{k}^{n+1}\right|_t}  \bm{f}_{t,k}^{n+1},
\end{eqnarray}
where $(\cdot)|_t$ denotes quantities evaluated at $\bm{x}_t$ using second order trilinear interpolations. The term $\bm{f}_{t,k}^{n+1}$ is computed in Step 2.2 of the prior sub-iteration.
       
\textbf{Step 1.2} Next, we update the particle centroids with a second order Adams-Moulton method,
\begin{eqnarray}
	\bm{x}_{p,k+1}^{n+1}&=&\bm{x}_{p}^{n} + \Delta t\; \bm{u}_{p,k+1}^{n+1}\\
	\bm{u}_{p,k+1}^{n+1}&=&\bm{u}_{p}^{n}+ \frac{\Delta t}{m_p}\left\{\frac{1}{2}\left(\bm{F}_{h,p,k}^{n+1}+\bm{F}_{h,p}^{n}\right) + m_p\bm{g}\right\}\\
	\bm{\omega}_{p,k+1}^{n+1}&=&\bm{\omega}_{p}^{n}+ \frac{\Delta t}{I_p}\left\{\frac{1}{2}\left(\bm{T}_{h,p,k}^{n+1}+\bm{T}_{h,p}^{n}\right)\right\}
\end{eqnarray}

\textbf{Step 1.3} In this step, we update the centroids of the surface elements on each particle. This amounts to solving the rigid body equations,
\begin{eqnarray}
	\frac{d}{dt}(\bm{x}_t-\bm{x}_p)&=&\bm{\omega}_p\times (\bm{x}_t-\bm{x}_p)\label{eq:rigid_body_1}\\
	\frac{d}{dt}\bm{n}_t&=&\bm{\omega}_p\times \bm{n}_t\label{eq:rigid_body_2}
\end{eqnarray}
for all surface element centroids $\bm{x}_t$ and normals $\bm{n}_t$ on a particle ``$p$''. Integrating equations (\ref{eq:rigid_body_1}) and (\ref{eq:rigid_body_2}) must be carried out with great care to avoid unphysical deformation of the particles. For this reason, we solve these equations exactly which yields the following updates,
\begin{eqnarray}
	\bm{x}_{t,k+1}^{n+1}&=& \bm{x}_{p,k+1}^{n+1} + \bm{R}_p(\Delta t) (\bm{x}_{t,k+1}^{n} - \bm{x}_p^{n}),\\
	\bm{n}_{t,k+1}^{n+1}&=& \bm{R}_p(\Delta t)\bm{n}_{t,k+1}^{n}.
\end{eqnarray}
Here, $\bm{R}_p(\Delta t)$ denotes a rotation matrix with axis $\bm{\omega}^{n+1}_{p,k+1}/\omega^{n+1}_{p,k+1}$, angle $\theta=\Delta t \omega^{n+1}_{p,k+1}$, and where $\omega^{n+1}_{p,k+1}=\sqrt{\bm{\omega}^{n+1}_{p,k+1}\cdot\bm{\omega}^{n+1}_{p,k+1}}$. The product between $\bm{R}_p(\Delta t)$ and an arbitrary vector $\bm{a}$ is easily evaluated using the formula

\begin{eqnarray}
	\bm{R}_p(\Delta t)\bm{a} &=& \cos(\theta)\bm{a}+\sin(\theta) \frac{\bm{\omega}^{n+1}_{p,k+1}}{\omega^{n+1}_{p,k+1}}\times\bm{a}+ \left(1-\cos(\theta)\right) \frac{(\bm{\omega}^{n+1}_{p,k+1}\cdot\bm{a})\ \bm{\omega}^{n+1}_{p,k+1}}{\bm{\omega}^{n+1}_{p,k+1}\cdot\bm{\omega}^{n+1}_{p,k+1}}.
\end{eqnarray}

\textbf{Step 1.4} Next, we update the surface density function $\Sigma$,
\begin{equation}
	\Sigma_{k+1}^{n+1}(\bm{x})= \sum_p \sum_{t\in S_p} A_t \mathcal{G}(\bm{x}-\bm{x}_{t,k+1}^{n+1})
\end{equation}
The procedure to build the filter kernel $\mathcal{G}(\bm{x}-\bm{x}_{t,k+1}^{n+1})$ on the fluid grid is described in detail in  \S3.3.2 of \citep{daveVolumefilteringImmersedBoundary2023}.

\textbf{Step 1.5} The final step of the particle solver consists in computing the new fluid volume fraction $\alpha_{f,k+1}^{n+1}$ by solving the  following Poisson equation:
\begin{equation}
	\nabla^2\alpha_{f,k+1}^{n+1}= \nabla\cdot \left(\sum_p \sum_{t\in S_p} \bm{n}_{t,k+1}^{n+1} A_t \mathcal{G}(\bm{x}_{t,k+1}^{n+1}-\bm{x}_{p,k+1}^{n+1})\right)
\end{equation}
Further details can be found in \citep{daveVolumefilteringImmersedBoundary2023}.

\paragraph{Step 2 (Fluid Solver)} In the next steps, we advance the fluid velocity and pressure to $n+1$ using a predictor-corrector approach.

\textbf{Step 2.1} We start by computing an intermediate flow field using a Crank-Nicolson method:
\begin{equation}
	\breve{\bm{u}}_{k+1}^{n+1}= \bm{u}^n + \Delta t \nabla p_{k}^{n+1} + \Delta t\mathcal{M}\left(\frac{1}{2}(\bm{u}_k^{n+1} + \bm{u}^n)\right)  + \Delta t \frac{\partial\mathcal{M}}{\partial \bm{u}}\left(\frac{1}{2}(\tilde{\bm{u}}_{k+1}^{n+1} + \bm{u}_k^{n+1})\right)
\end{equation}
where $\mathcal{M}$ represents the momentum operator,
\begin{equation}
	\mathcal{M}(\bm{u})=-\nabla\cdot (\bm{u}\bm{u}) +\frac{\mu}{\rho}\nabla^2\bm{u}.
\end{equation}
The Jacobin $\partial \mathcal{M}/\partial\bm{u}$ allows the treatment of the non-linearity with a Newton-Raphson method \citep{akselvollLargeEddySimulation1995,pierceProgressVariableApproachLargeEddy2001}.
      
\textbf{Step 2.2} Next, we compute a second intermediate velocity that accounts for the interphase-momentum coupling term,
\begin{equation}
	\tilde{\bm{u}}_{k+1}^{n+1} = \breve{\bm{u}}_{k+1}^{n+1} + \Delta t \bm{F}_{I,k+1}^{n+1}
\end{equation}
where the last term is discretized as
\begin{equation}
	\bm{F}_{I,k+1}^{n+1}=\sum_{p} \sum_{t\in S_p} \left(\frac{A_t}{\left.\Sigma_{k}^{n+1}\right|_t} \bm{f}_{t,k+1}^{n+1} \mathcal{G}(\bm{x}_{t,k+1}^{n+1}-\bm{x}_{p,k+1}^{n+1})\right)
\end{equation}
and the forces $\bm{f}_{t,k+1}^{n+1}$ on the centroids of the surface elements are 
\begin{equation}
	\bm{f}_{t,k+1}^{n+1}= \rho_f\frac{\bm{u}_{t,k+1}^{n+1}-\left. \breve{\bm{u}}_{k+1}^{n+1}\right|_t}{\Delta t}.
\end{equation}
Note that, rigid body dynamics dictate that the velocities $\bm{u}_{t,k+1}^{n+1}$ at the centroid $\bm{x}_t$ of triangular surface elements are 
\begin{equation}
	\bm{u}_{t,k+1}^{n+1}= \bm{u}_{p,k+1}^{n+1}+\bm{\omega}_{p,k+1}^{n+1}\times\left(\bm{x}_{t,k+1}^{n+1}-\bm{x}_{p,k+1}^{n+1}\right).
\end{equation}

\textbf{Step 2.2} Next, we find the pressure at sub-iteration $k+1$ and time-level $n+1$ by solving the pressure-Poisson equation for $\phi=p_{k+1}^{n+1}-p_k^{n+1}$,
\begin{eqnarray}
	\nabla^2\phi=\frac{\rho_f}{\Delta t} \nabla \cdot \hat{\bm{u}}_{k+1}^{n+1}
\end{eqnarray}

\textbf{Step 2.3} Finally, we apply the pressure correction to get a divergence-free solution:
\begin{eqnarray}
	  p_{k+1}^{n+1} &=& p_k^{n+1} + \phi\\
	  \bm{u}_{f,k+1}^{n+1}&=&\hat{\bm{u}}_{f,k+1}^{n+1}-\frac{\Delta t}{\rho_f}\nabla \phi
\end{eqnarray}

\paragraph{Step 3}: Repeat steps 1-2 until $k=k_\mathrm{max}$.

Although, in principle, the iterative procedure should be carried for a large number of sub-iterations \citep{akselvollLargeEddySimulation1995,pierceProgressVariableApproachLargeEddy2001}, we have found that $k_\mathrm{max}=2$ is sufficient to yield accurate results as we show next.

%% file: sections/role_filter.tex
\subsection{Flow past a sphere and the role of sub-filter stresses}\label{ssec:role_filter}
\begin{figure}\centering
  \begin{subfigure}{0.49\linewidth}
    \includegraphics[width=3.2in]{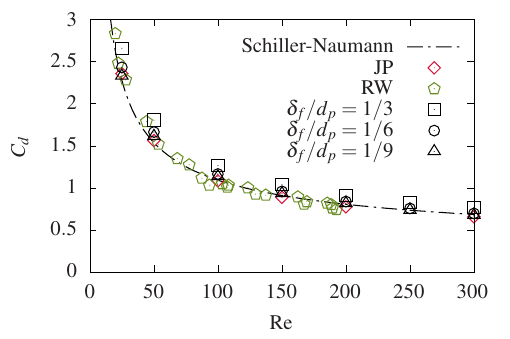}  
    \caption{\label{fig:filter_2_a}}
  \end{subfigure} 
  \begin{subfigure}{0.49\linewidth}
    \includegraphics[width=3.2in]{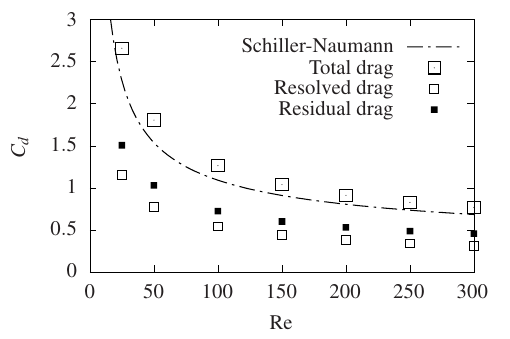}  
    \caption{\label{fig:filter_2_b}}
  \end{subfigure}
  \begin{subfigure}{0.49\linewidth}
    \includegraphics[width=3.2in]{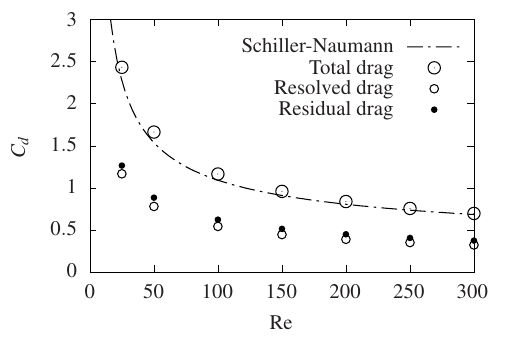}  
    \caption{\label{fig:filter_2_c}}
  \end{subfigure}
  \begin{subfigure}{0.49\linewidth}
    \includegraphics[width=3.2in]{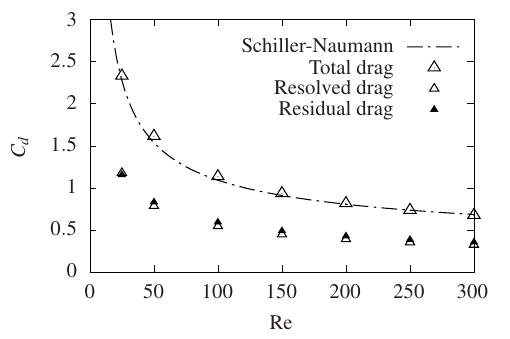}  
    \caption{\label{fig:filter_2_d}}
  \end{subfigure}
  \caption{Variation of the drag coefficient with Reynolds number in PR-DNS with a fixed sphere. Figure \ref{fig:filter_2_a} shows the predicted drag converges towards the Schiller-Naumann correlation with improving resolution of the particle. Figures \ref{fig:filter_2_b}--\ref{fig:filter_2_d} shows the relative contributions of the resolved and residual drag at the resolution $\delta_f/d_p=1/3$, 1/6, and 1/9, respectively. \label{fig:filter_2}}
\end{figure}	

In this section, we discuss the relative importance of resolved and residual stresses on the hydrodynamic force exerted on a particle. For this purpose, we consider the flow past a sphere of diameter $d_p$ at Reynolds number $\Rey=U d_p/\nu=25$ up to 300, where $U$ is the upstream velocity. The numerical configuration is identical to the one used in \citep{daveVolumefilteringImmersedBoundary2023}. The computational domain is long by $26d_p$ in the flow direction, and by $16 d_p$ in the other two directions. The sphere is located at a distance $6 d_p$ from the inlet. 

We perform simulations at three resolution levels: a coarse resolution with $\delta_f=(1/3)d_p$, an intermediate resolution with $\delta_f=(1/6)d_p$, and a fine resolution with $\delta_f=(1/9)d_p$. In all cases, we ensure that the filter kernel is resolved with 4 grid points, i.e., $\Delta x=\delta_f/4$. This is sufficient to converge the results with respect to $\Delta x$ \citep{daveVolumefilteringImmersedBoundary2023}. In terms of the ratio $d_p/\Delta x$, we get 12, 24, and 36 for the coarse, intermediate and fine resolutions, respectively. In addition, we use a Delaunay triangulation with typical element size $\Delta_s\sim \delta_f/8$ to discretize the surface of the particle. We carry the integration until $20\times d_p/U$, by which point the flow becomes steady.

Figure \ref{fig:filter_2_a} shows the predicted drag coefficient with the present PR-DNS strategy for Reynolds numbers between $\Rey_D=25$ and $300$.  The numerical results compare very well with the Schiller-Naumann correlation $C_D=(24/\Rey)\times(1+0.15\,\Rey^{0.687})$, the body-fitted simulations of \citet{johnsonFlowSphereReynolds1999} (JP), and the experiments of  \citet{roosExperimentalResultsSphere1971} (RW) for the full range of Reynolds numbers considered. The agreement is good even at the lowest resolution $\delta_f=(1/3)d_p$, yielding values of the drag that are within 15\% of the Schiller-Naumann correlation. This agreement improves with increasing resolution such that the deviations from the Schiller-Naumann correlation drop to within 2\% at $\delta_f=(1/9)d_p$. 

The contributions of the resolved drag and residual drag are shown in figures \ref{fig:filter_2_b}, \ref{fig:filter_2_c}, and \ref{fig:filter_2_d} for the cases with coarse, intermediate, and fine resolution, respectively. In general, the drag due to residual stresses exceeds the drag resulting from the filtered stresses. Based on the data in figures \ref{fig:filter_2_b}--\ref{fig:filter_2_d}, the Reynolds number does not seem to have an impact on the relative contributions of the resolved and residual drag. However, as the ratio $\delta_f/d_p$ decreases, which indicates better resolution of the surface, the discrepancy between the relative contributions of the resolved and filtered drag reduces. Figure \ref{fig:filter_2_d} suggests that in the limit $\delta_f/d_p\ll 1$, both resolved and residual drag have equal contributions.


%% file: sections/unbounded_settling.tex
\subsection{Unbounded settling}\label{ssec:unbounded_settling}

\begin{figure}\centering
	\begin{subfigure}{\linewidth}\centering
	  \includegraphics[width=3.8in]{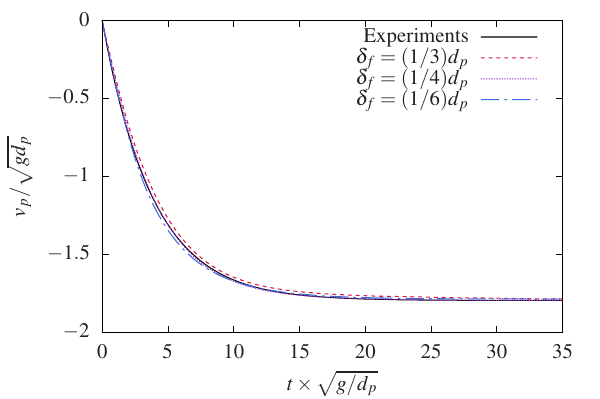}
	  \caption{}	
	\end{subfigure}
	\begin{subfigure}{\linewidth}\centering
	  \includegraphics[width=3.8in]{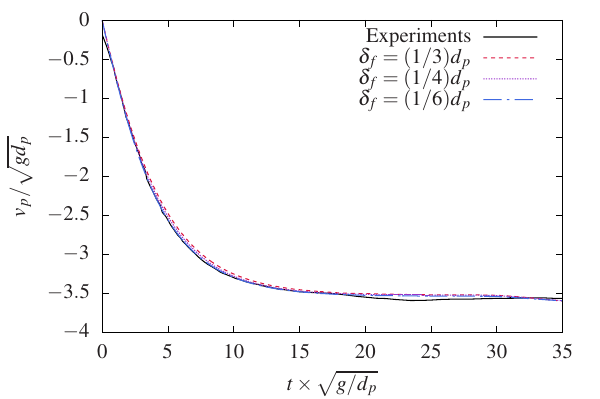}
	  \caption{}	
	\end{subfigure}
	\begin{subfigure}{\linewidth}\centering
	  \includegraphics[width=3.8in]{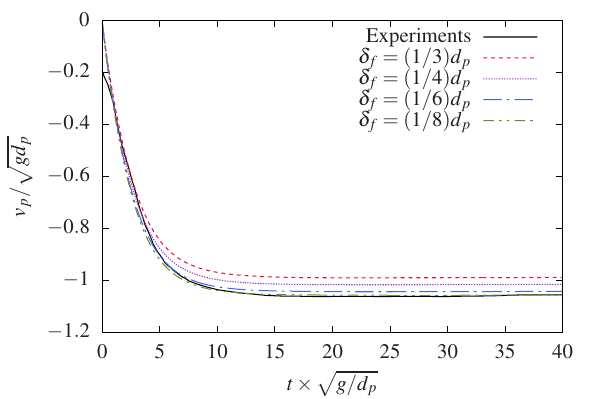}
	  \caption{}	
	\end{subfigure}
	\caption{Evolution of the settling velocity of a single particle in an unbounded domain at (a) $\rho_p/\rho_f=2.56$ and $\mathrm{Ga}=255.35$, (b) $\rho_p/\rho_f=7.71$ and $\mathrm{Ga}=206.17$, and (c) $\rho_p/\rho_f=2.56$ and $\mathrm{Ga}=49.12$. Dashed line correspond to simulations with the present PR-DNS strategy at varying filter size. The solid black lines correspond to the experimental data from \citet{mordantVelocityMeasurementSettling2000}. \label{fig:settling}}
\end{figure}

We now consider the settling of a single particle in an initially quiescent flow. The present simulations are the numerical analogous of the experiments carried out by \citet{mordantVelocityMeasurementSettling2000}. Two non-dimensional numbers control the settling behavior, the density ratio $\rho_p/\rho_f$ and the Galileo number   $\mathrm{Ga}=\sqrt{(\rho_p/\rho_f-1)gd_p^3}/\nu$.
For this comparison, we focus on three well-characterized cases in \citep{mordantVelocityMeasurementSettling2000} which yield the pairs $(\rho_p/\rho=2.56, \mathrm{Ga}=255.35)$, $(\rho_p/\rho=7.71, \mathrm{Ga}=206.17)$, and $(\rho_p/\rho=2.56, \mathrm{Ga}=49.12)$.

We perform simulations in a fully periodic domain with gravity applied along the $y$-direction. In the horizontal plane, the domain has dimensions $8d_p$-by-$8d_p$. In the vertical direction, the domain is long by $60d_p$ for the cases at $\mathrm{Ga}=255.35$ and $\mathrm{Ga}=49.12$ and by $90 d_p$ for the case at $\mathrm{Ga}=206.17$. The discretization is uniform and depends on the filter size $\delta_f$ as we maintain $\delta_f/\Delta x=4$ in all cases. The coarsest filter size we consider is $\delta_f=(1/3)d_p$, which yields 12 grid points across the diameter and grid sizes $96\times 720\times 96$ and $96\times 1080\times 96$. The simulation with finest filter has $\delta_f=(1/8)d_p$, resulting in 32 grid points across the diameter and grid sizes $256\times 1920\times 256$ and $256\times 2880\times 256$. The other filter sizes we consider are $\delta_f=(1/4)d_p$ and $\delta_f=(1/6)d_p$ which yield 16 and 24 grid points across the diameter, respectively. We also apply a forcing that maintains a constant mass flow rate in the vertical direction. This forcing represents a Galilean invariant transformation \citep{tennetiDragLawMonodisperse2011} and is necessary since in our formulation gravity applies explicitly in the governing equations of both fluid and solid phases. All simulations are carried out until $t\times\sqrt{g/d_p}=35$ .

Figure \ref{fig:settling} shows time series of the normalized vertical particle velocity $v_p/\sqrt{g d_p}$  for the three cases at $(\rho_p/\rho=2.56, \mathrm{Ga}=255.35)$, $(\rho_p/\rho=7.71, \mathrm{Ga}=206.17)$, and $(\rho_p/\rho=2.56, \mathrm{Ga}=49.12)$. The agreement between PR-DNS and experiments of \citet{mordantVelocityMeasurementSettling2000} is excellent both during the acceleration phase and the steady state, even at the lowest filter size $\delta_f=(1/3)d_p$. The agreement improves with decreasing $\delta_f$, which also demonstrate the convergence of the method under filter size refinement.

\begin{table}
	\caption{Comparison of the Froude number $\mathrm{Fr}=v_{p,\infty}/\sqrt{g d_p}$, Reynolds number $\Rey=v_{p,\infty} d_p/\nu$, and normalized characteristic time $\tau_{95}\sqrt{g/d_p}$ from the present simulations and the experiments of \citet{mordantVelocityMeasurementSettling2000}.\label{tab:settling}}
	\newcolumntype{b}{X}
	\newcolumntype{s}{>{\hsize=.50\hsize}X}
	\begin{tabularx}{\linewidth}{bsssssss}\hline
		& $\mathrm{Ga}$ & $\rho_p/\rho_f$ & $\delta_f/d_p$ & $d_p/\Delta x$ & $v_{p,\infty}/\sqrt{g d_p}$ & $v_{p,\infty} d_p/\nu$ & $\tau_{95}\sqrt{g/d_p}$ \\\hline
		Present  		& 255.35 & 2.56 & 1/3 & 12 & 1.79 & 365.35 & 11.85 \\
		Present  		&        &      & 1/4 & 16 & 1.80 & 367.44 & 11.18 \\
		Present  		&        &      & 1/6 & 24 & 1.79 & 365.47 & 10.97 \\
		Experiments \citep{mordantVelocityMeasurementSettling2000} &        &      & -- & -- & 1.80 & 367.41 & 11.48 \\ \hline
		Present  		& 206.17 & 7.71 & 1/3 & 12 & 3.60 & 286.61 & 13.48 \\
		Present  		&        &      & 1/4 & 16 & 3.60 & 286.32 & 12.75 \\ 
		Present  		&        &      & 1/6 & 24 & 3.60 & 286.39 & 12.44 \\
		Experiments \citep{mordantVelocityMeasurementSettling2000} &        &      & -- & -- & 3.57 & 284.04 & 11.96 \\ \hline
		Present  		& 49.12  & 2.56 & 1/3 & 12 & 0.99 & 38.96 & 7.68 \\
		Present  		&        &      & 1/4 & 16 & 1.02 & 40.01 & 7.49 \\
		Present  		&        &      & 1/6 & 24 & 1.04 & 41.07 & 7.37 \\
		Present  		&        &      & 1/8 & 32 & 1.06 & 41.60 & 7.27 \\
		Experiments \citep{mordantVelocityMeasurementSettling2000} &        &      & -- & -- & 1.06 &  41.63 & 7.70 \\ \hline
	\end{tabularx}	
\end{table}

For a more quantitative comparison, we report in table \ref{tab:settling} the normalized terminal velocity $v_{p,\infty}/\sqrt{gd_p}$ (which is also equal to the Froude number $\mathrm{Fr}$), terminal Reynolds number $\Rey= v_{p,\infty} d_p/\nu$, and normalized time taken by the particle to reach 95\% of its terminal velocity $\tau_{95}\sqrt{g/d_p}$. At $\delta_f=(1/3)d_p$, the difference in terminal settling velocity between the experiments and simulations is about 0.6\%, 0.9\%, and 6.4\% for the cases at $\mathrm{Ga}=255.35$, 206.17, and 49.12, respectively. At $\delta_f=(1/6)d_p$, the discrepancy between experiments and simulations drops to 0.5\%, 0.8\%, and 1.3\% in these three cases. For the case at $\mathrm{Ga}=49.12$, further improving the resolution with $\delta_f=(1/8)d_p$ reduces the discrepancy in settling velocity to about 0.07\%. The characteristic times $\tau_95$ computed from the present simulations also agree very well with the experimental values as shown in table \ref{tab:settling}.

If one considers the simulations to be converged when the terminal velocity is predicted within 5\% of the reference values, then this is achieved with the present strategy for relatively coarse simulations: with $\delta_f=(1/3)d_p$ (i.e., $d_p/\Delta x=12$) for the cases at $\mathrm{Ga}=255.35$ and 206.17, and $\delta_f=(1/4)d_p$ (i.e., $d_p/\Delta x=16$) for the case at $\mathrm{Ga}=49.12$.

%% file: sections/low_density.tex
\subsection{Low density ratio and neutrally buoyant particles}\label{ssec:low_density}

To show the robustness of the force calculation using equation (\ref{eq:part_6}) for particles with low density ratios, we revisit the case of a freely settling particle from the previous section.  We consider 6 cases where $\rho_p/\rho_f$ is progressively reduced from 2.56 to 1.0. The case at $\rho_p/\rho_f=2.56$ is the one previously analyzed in \S\ref{ssec:unbounded_settling}, and for which $\mathrm{Ga}=49.12$. The case where $\rho_p/\rho_f=1$ corresponds to a neutrally buoyant particle. Save for the varying particle density, the particle diameter, fluid density, viscosity and gravity are identical in the all these cases. The numerical parameters are also identical in all cases, with $\delta_f=(1/4)d_p$.

\begin{figure}\centering
	\begin{tikzpicture}
		\node (fig) at (0,0) {\includegraphics[width=5in]{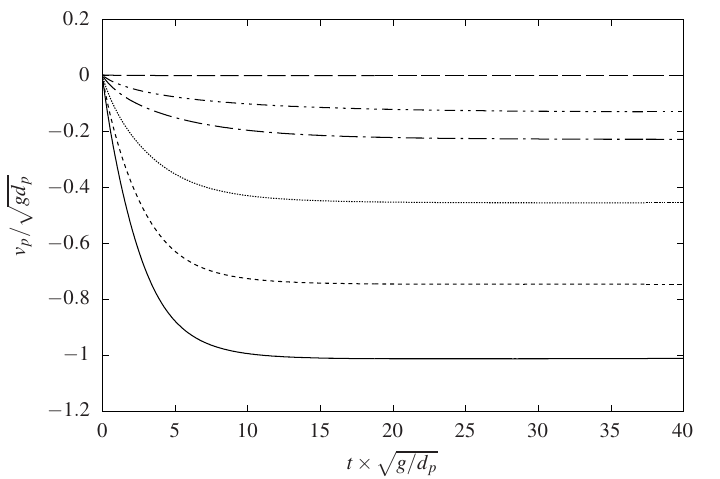}};
		\begin{scope}[shift={(fig.south west)},x=(fig.south east), y=(fig.north west)]
			\node (R256) at (0.35,0.22) {\small $\rho_p/\rho_f=2.56$};
			\node (R200) at (0.42,0.38) {\small $\rho_p/\rho_f=2.0$};
			\node (R150) at (0.50,0.54) {\small $\rho_p/\rho_f=1.5$};
			\node (R120) at (0.58,0.67) {\small $\rho_p/\rho_f=1.2$};
			\node (R110) at (0.68,0.89) {\small $\rho_p/\rho_f=1.1$};
			\node (R100) at (0.32,0.90) {\small $\rho_p/\rho_f=1.0$};
			\draw (R110) -- (0.72,0.76);
			\draw (R100) -- (0.36,0.83);
		\end{scope}
	\end{tikzpicture}
	
	\caption{Particle settling velocity at low density ratios. \label{fig:settling_low_density_ratio}}
\end{figure}
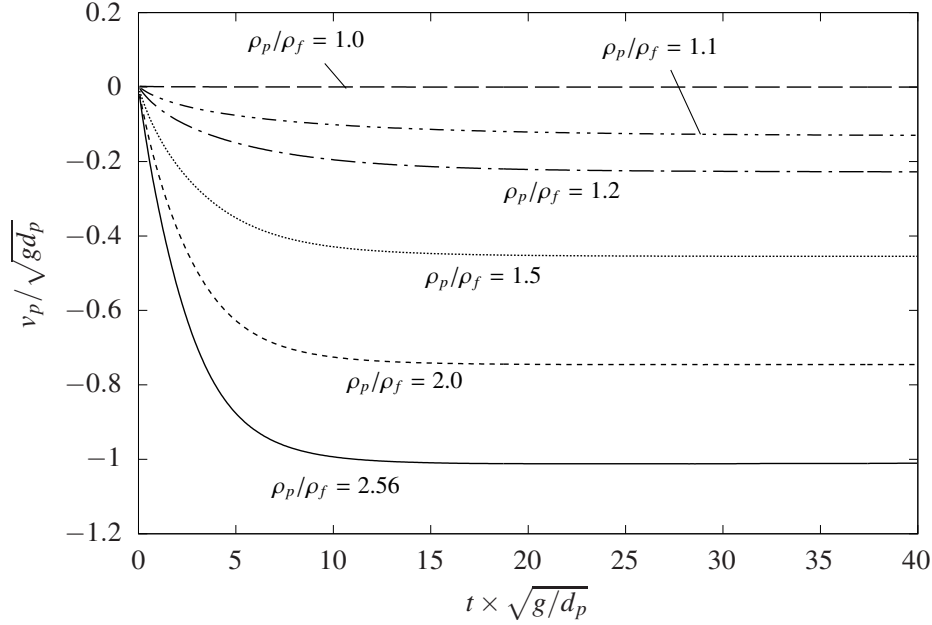

Figure \ref{fig:settling_low_density_ratio} shows a time series of the particle vertical velocity. All runs converge and are stable. This includes cases where $\rho_p/\rho_f<1.2$, which are known to diverge using force calculation methods based on integrating the flow inside the particle \citep{uhlmannImmersedBoundaryMethod2005}.

As one expects, figure \ref{fig:settling_low_density_ratio} shows that the particle terminal velocity reduces with $\rho_p/\rho_f$. In the case $\rho_p/\rho_f=1$,  the particle vertical velocity remains identically zero during the entire run, as one would expect for a neutrally buoyant particle. In this case, the buoyancy force and gravity balance exactly preventing any motion of the particle.

%% file: sections/confined_settling.tex
\subsection{Settling under confinement}\label{ssec:confined}

For a more quantitative validation of the approach at low density ratios, we reproduce numerically the experiments of \citet{tencateParticleImagingVelocimetry2002}. In the latter, a 15mm sphere with density 1120$\mathrm{kg/m^3}$ is placed in a box of size 100mm$\times$160mm$\times$100mm filled with a viscous fluid. 
\citet{tencateParticleImagingVelocimetry2002} consider four cases where the fluid  density and viscosity are varied to yield $(\rho_f=970\,\mathrm{kg/m^3},\mu_f=373\times 10^{-3}\,\mathrm{Ns/m^2})$, $(\rho_f=965\,\mathrm{kg/m^3},\mu_f=212\times 10^{-3}\,\mathrm{Ns/m^2})$, $(\rho_f=962\,\mathrm{kg/m^3},\mu_f=113\times 10^{-3}\,\mathrm{Ns/m^2})$, and $(\rho_f=960\,\mathrm{kg/m^3},\mu_f=58\times 10^{-3}\,\mathrm{Ns/m^2})$. In terms of non-dimensional number, these cases correspond to the pairs $(\rho_p/\rho=1.155, \mathrm{Ga}=5.88)$, $(\rho_p/\rho=1.160, \mathrm{Ga}=10.55)$, $(\rho_p/\rho=1.164, \mathrm{Ga}=19.85)$, and $(\rho_p/\rho=1.167, \mathrm{Ga}=26.53)$. While the density ratio $\rho_p/\rho_f\sim 1.1$ does not vary significantly in these four cases, it is the variation of the Galileo number that leads to faster or slower particle settling.

\begin{figure}\centering
	\includegraphics[width=4.5in]{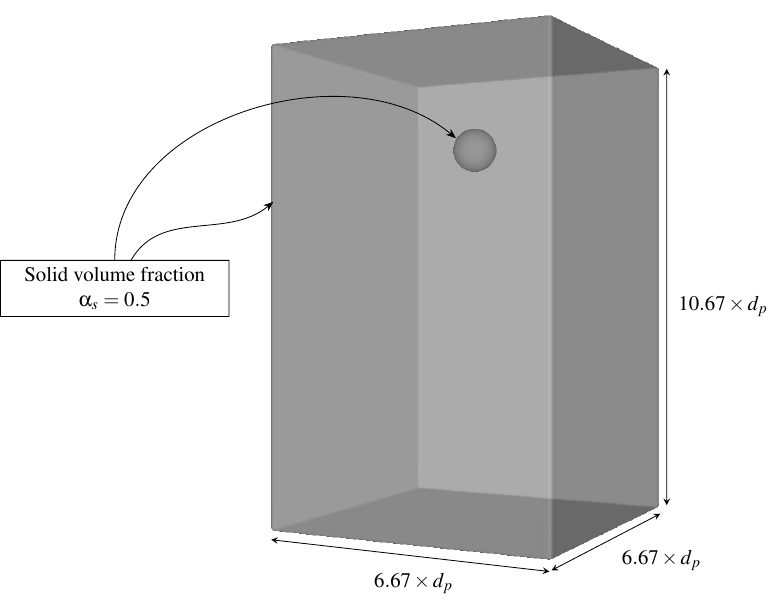}
	\caption{Configuration for the comparison with the experiments of \citet{tencateParticleImagingVelocimetry2002} showing the initial location of the sphere at height $y_p=(8.5)d_p$ from the bottom. The gray transparent surface corresponds to the isocontour $\alpha_s=0.5$. Both particle and enclosure walls are represented with the same volume-filtering immersed boundary method.
	\label{fig:tenCate_config}}
\end{figure}

\begin{figure}\centering
	\begin{subfigure}{\linewidth}\centering
		\includegraphics[width=\linewidth]{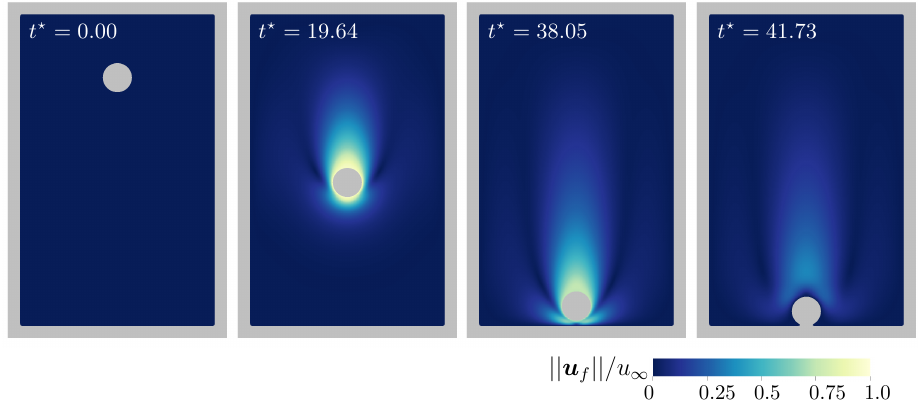}
		\caption{\label{fig:tenCate_a}}
	\end{subfigure}\\
	\begin{subfigure}{\linewidth}
		\includegraphics[width=\linewidth]{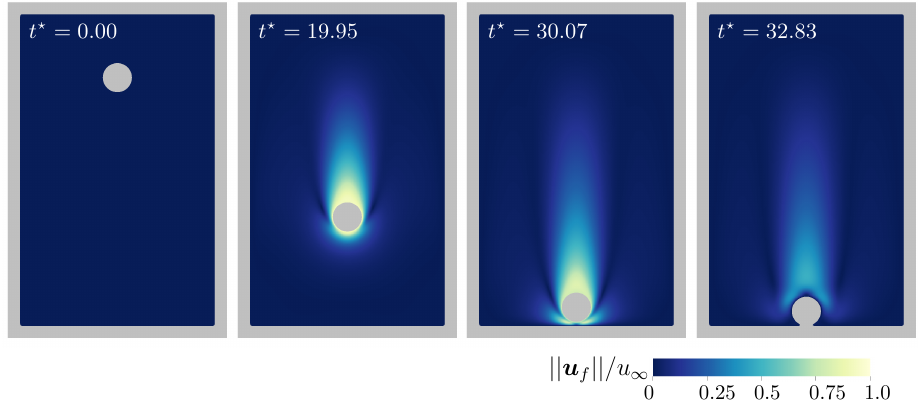}
		\caption{\label{fig:tenCate_b}}
	\end{subfigure}
	\caption{Normalized velocity magnitude at different non-dimensional times $t^\star=t\times\sqrt{g/d_p}$ for the case at (a) $(\rho_p/\rho=1.164, \mathrm{Ga}=19.85)$ and (b) $(\rho_p/\rho=1.167, \mathrm{Ga}=26.53)$. The gray regions are those where $\alpha_s>0.5$, thus corresponding to the solid phase. \label{fig:tenCate_vmag}}
\end{figure}

Figure \ref{fig:tenCate_config} shows the enclosure and sphere upon initialization in our numerical analogue. Similar to the experiments,  the enclosure has dimensions $6.67d_p\times 10.67d_p\times 6.67d_p$  and the sphere is initially located at the vertical location $y_p=8.5d_p$ from the bottom. Note that both sphere and enclosure walls are all represented with the volume-filtering immersed boundary method presented here and in our prior study \citep{daveVolumefilteringImmersedBoundary2023}. 
	The simulation domain is $7.5d_p\times 11.5d_p\times 7.5d_p$, which is slightly larger than enclosure, which allows us to compute the solid volume fraction within the solid enclosure as well as within the particle using equation (\ref{eq:volume_fraction}). In these simulations, the resolution is $\delta_f=(1/6)d_p$, which yields 24 grid points across the particle diameter. 

Figure \ref{fig:tenCate_vmag} shows the settling of the sphere for the cases at $\mathrm{Ga}=19.85$ and $\mathrm{Ga}=26.53$. After a short transient, the spheres reach their terminal settling velocity. Once the particles are within $\sim d_p$ from the bottom wall, they start decelerating until collision occurs, which we handle using a basic soft-sphere collision model, namely, the linear spring-dashpot model \citep{cundallDiscreteNumericalModel1979}.

\begin{figure}\centering
	\begin{subfigure}{.49\linewidth}\centering
		\includegraphics[width=3.2in]{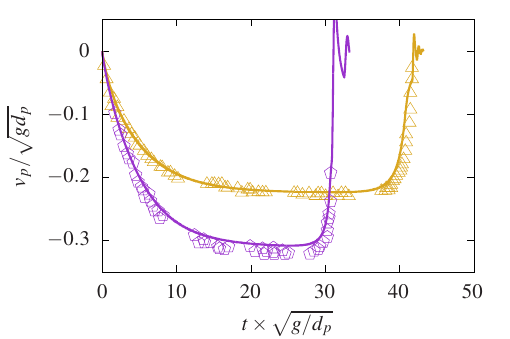}
		\caption{\label{fig:tenCate_comparison_a}}
	\end{subfigure}\hfill
	\begin{subfigure}{.49\linewidth}
		\includegraphics[width=3.2in]{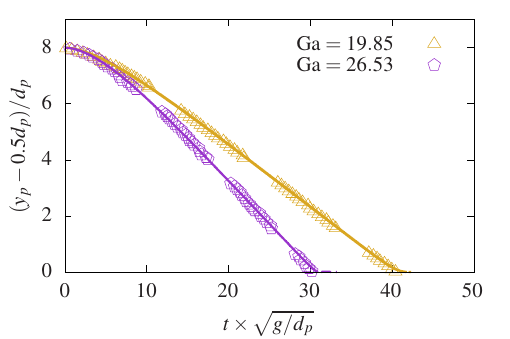}
		\caption{\label{fig:tenCate_comparison_b}}
	\end{subfigure}
	\begin{subfigure}{.49\linewidth}\centering
		\includegraphics[width=3.2in]{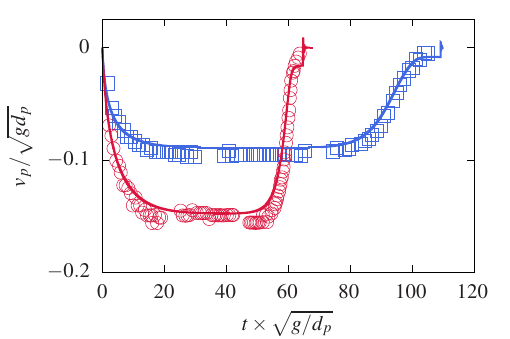}
		\caption{\label{fig:tenCate_comparison_a}}
	\end{subfigure}\hfill
	\begin{subfigure}{.49\linewidth}
		\includegraphics[width=3.2in]{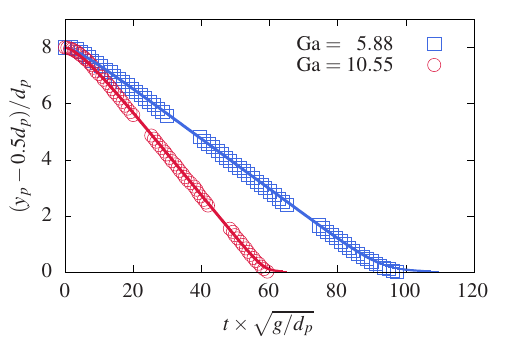}
		\caption{\label{fig:tenCate_comparison_b}}
	\end{subfigure}
	\caption{Evolution of the (a,c) settling velocity and (b,d) normalized particle position in the present simulations (lines) and the experiments of \citet{tencateParticleImagingVelocimetry2002} (symbols). The simulations are carried out with a filter size $\delta_f=(1/6)d_p$. \label{fig:tenCate_comparison}}
\end{figure}

Figure \ref{fig:tenCate_comparison} shows a comparison of the particle settling velocity and normalized height from the present simulations with $\delta_f=(1/8)d_p$ and the experiments of \citep{tencateParticleImagingVelocimetry2002}. As one can see from the figures, the agreement is very good throughout the entire trajectory of the particle and in all four cases. This demonstrates the capacity of the present strategy to capture the dynamics of particle-laden flows at low density ratios with high-fidelity.

%% file: sections/conclusion.tex
\section{Conclusion}\label{sec:conclusion}
We developed a computational strategy for high-fidelity simulations of flows laden with fully resolved particles. This approach is robust, stable, and reproduces experimental benchmarks with high accuracy for a wide range of particle-to-fluid density ratios. In particular, the method is stable for low-density ratio and neutrally buoyant particles without any special treatment.

To build the mathematical framework of this method, we used the well-established volume-filtering method \citep{andersonFluidMechanicalDescription1967} to derive filtered equations, which we then integrate numerically. This modeling approach is well-established in the multiphase flow community, and has been previously used to derive Euler-Lagrange models for particle-laden flows. In those prior studies, the filter width $\delta_f$ is assumed to be much larger than the particle size $d_p$ to justify the  point-particle view of the disperse solid phase. In the present work, we adopt the opposite limit $\delta_f\ll d_p$. This makes the particle geometry, boundary layer surrounding the particle, and all other flow features well-resolved by the filter. This approach requires no closures and offers the highest possible fidelity for simulations of particle-laden flows.

Our method shows excellent agreements with benchmark experimental data for cases with fixed and moving particles. In our tests, a coarse filter $\delta_f=(1/3)d_p$ is often times sufficient to get the drag force within 10\% of the reference values. Since we ensure that $\Delta x=\delta_f/4$, this coarse filter yields 12 grid points across the particle diameter. The agreement with the reference values falls within 1 or 2\% when using a fine filter, such as $\delta_f=(1/8)d_p$ which yields 32 grid points across the particle diameter. We verified these results in benchmark cases of flow past a fixed sphere at various Reynolds numbers, settling spheres in an unbounded medium, and settling spheres in a confined box at density ratios $\rho_p/\rho_f\sim 1.1$ and different Galileo numbers .

We also showed the importance of the filtered and residual drag.  These two contributions have not been considered in prior PR-DNS methods, but arise organically in the volume-filtering framework. We showed that the total drag exerted on a particle has two contributions originating from  the filtered fluid stress tensor $\overline{\bm{\tau}}$ and from the residual tensor $\bm{\tau}'=\bm{\tau}-\overline{\bm{\tau}}$. When $\delta_f/d_p$ is vanishingly small, the residual drag accounts for at least 50\% of the total drag. For small but intermediate filter sizes such as $\delta_f=(1/3)d_p$ or $\delta_f=(1/4)d_p$, the residual drag is greater than the filtered drag, although the two contributions have similar order of magnitude. This is to be contrasted with Euler-Lagrange methods, i.e., the limit where $\delta_f\gg d_p$, where the total drag is often assumed to be completely dominated by the residual drag.